\documentclass[sigconf]{acmart}

\AtBeginDocument{%
  }

\copyrightyear{2023}
\acmYear{2023}
\setcopyright{acmcopyright}
\acmConference[WSDM '23] {Proceedings of the Sixteenth ACM International Conference on Web Search and Data Mining}{February 27--March 3, 2023}{Singapore, Singapore.}
\acmBooktitle{Proceedings of the Sixteenth ACM International Conference on Web Search and Data Mining (WSDM '23), February 27--March 3, 2023, Singapore, Singapore}
\acmPrice{15.00}
\acmISBN{978-1-4503-9407-9/23/02}
\acmDOI{10.1145/XXXXXXX.XXXXXXX}

\usepackage{multirow}
\usepackage{bm}
\usepackage{bbm}
\usepackage[ruled,linesnumbered]{algorithm2e}
\usepackage{subfigure}
\usepackage{graphicx}
\usepackage{bbding}
\usepackage{balance}

\begin{document}

\title{Mining User-aware Multi-relations for Fake News Detection in Large Scale Online Social Networks}

\author{Xing Su, Jian Yang, Jia Wu, Yuchen Zhang}
\affiliation{%
\institution{School of Computing, Macquarie University}
\streetaddress{}
\city{Sdyney}
\state{New South Wales}
\country{Australia}
}
\email{{xing.su2, yuchen.zhang3}@students.mq.edu.au}
\email{{jian.yang, jia.wu}@mq.edu.au}




\renewcommand{\shortauthors}{Su et al.}

\begin{abstract}
Users' involvement in creating and propagating news is a vital aspect of fake news detection in online social networks. Intuitively, credible users are more likely to share trustworthy news, while untrusted users have a higher probability of spreading untrustworthy news. In this paper, we construct a dual-layer graph (\textit{i.e.}, the news layer and the user layer) to extract multi-relations of news and users in social networks to derive rich information for detecting fake news. Based on the dual-layer graph, we propose a fake news detection model \texttt{Us-DeFake}. It learns the propagation features of news in news layer and the interaction features of users in user layer. Through the inter-layer in graph, \texttt{Us-DeFake} fuses the user signals that contain credibility information into the news features, to provide distinctive user-aware embeddings of news for fake news detection. The training process conducts on multiple dual-layer subgraphs obtained by a graph sampler to scale \texttt{Us-DeFake} in large scale social networks. Extensive experiments on real-world datasets illustrate the superiority of \texttt{Us-DeFake} which outperforms all baselines, and the users' credibility signals learned by interaction relation can notably improve the performance of our model \footnote{Code is available at \url{https://github.com/xingsumq/Us-DeFake}}. 
\end{abstract}




\maketitle

\section{Introduction}
Nowadays people are accustomed to obtaining information on online social media, such as Twitter and Facebook. This convenient and quick access to information also brings a negative side, \textit{i.e.}, the brazen spread of fake news. For instance, the spread of COVID-19 fake news that ``consuming highly concentrated alcohol could disinfect the body and kill the virus'' resulted in more than 5,800 hospitalizations \cite{islam2020covid}. Fake news has caused severe harm to people's life \cite{van2020inoculating}.  Algorithms and mechanisms for fake news detection in online social networks are in urgent need for providing trustworthy information to the public.  

In order to detect fake news, the classic text-based methods \cite{pelrine2021surprising} detect fake news from the perspective of textual content of news, and distinguish linguistic features between real and fake news by capturing different writing styles \cite{guo2020future}. Recently, the development of graph neural networks (GNNs) \cite{wu2020comprehensive} shows promising results in fake news detection. Their capability of handling structural information allows them to explore the logical structure in news sentences or the news property of propagation. For graph-based methods, researchers generally construct graphs of words or sentences to capture textual structure \cite{yao2019graph}, or news propagation graphs to explore the structure of spreading \cite{ma-etal-2018-rumor}. Furthermore, some graph-based methods consider social contexts, \textit{e.g.}, constructing heterogeneous networks of news and users \cite{huang2020heterogeneous}, to provide more information for fake news detection \cite{weietal2021towards}. 

\begin{figure}[!tbp]
\centering
\centerline{\includegraphics[scale=0.09]{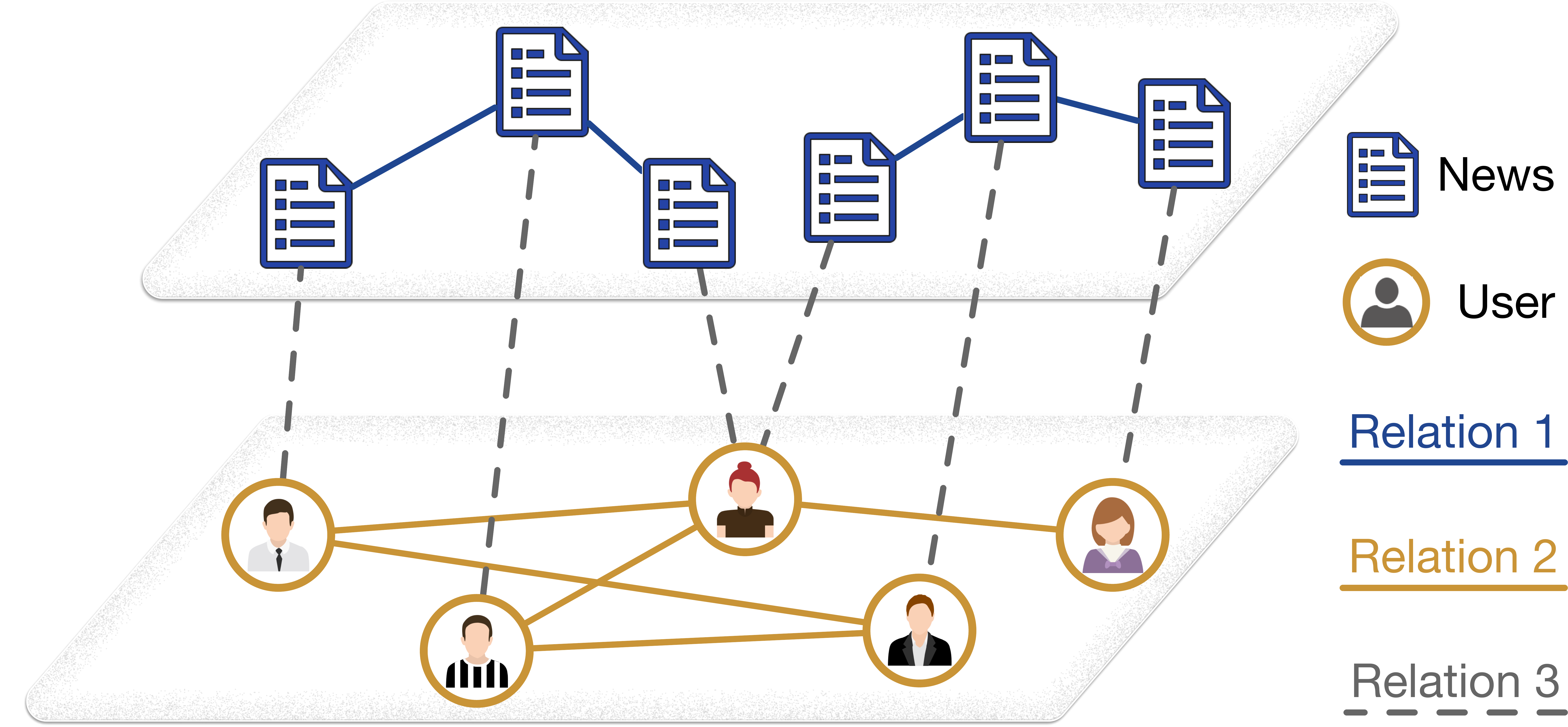}} 
\caption{An illustrative example of multiple relations in an online social network for fake news detection. \emph{Relation 1} represents the propagation relation of news, \emph{Relation 2} represents the interaction relation of users, and \emph{Relation 3} represents the posting relation between users and news.}
\label{fig:toy}
\end{figure}

Although the relation between news and users has been considered in some work, the interaction relation of users that causes fake news spread has not been fully explored \cite{nguyen2020fang}. As shown in Figure \ref{fig:toy}, there are essentially three relations in an online social network, \textit{i.e.,} the propagation relation of news, the interaction relation of users, and the posting relation between users and news. These relations are interdependent. The interaction patterns between users reflect the credibility of users. Reliable users are unlikely to spread fake news to their friends \cite{li2020let}. Therefore, user credibility reflected in user interaction can provide complementary information for fake news detection. In this space, existing methods still face the following challenges: (1) many methods only model part of relations for fake news detection and ignore the influence of user interaction on fake news spreading; (2) partial relations lead to inadequate information and mediocre performance in fake news detection; and (3) techniques for efficient detection in large scale social networks still require further development and improvement. 

To address the aforementioned challenges, we propose a method to capture \textbf{\underline{U}}ser-aware multi-relation\textbf{\underline{s}} for \textbf{\underline{De}}tecting \textbf{\underline{Fake}} news (\texttt{Us-DeFake}). As the graph in Figure \ref{fig:toy}, \texttt{Us-DeFake} regards news and users as nodes, constructs an attributed dual-layer network with intra-layers (\textit{i.e.}, news layer, user layer) and inter-layer (\textit{i.e.}, posting layer). This dual-layer graph contains the propagation relation of news, the interaction relation of users, and the posting relations between users and news. To capture information from these multi-relations to assist in fake news detection, the proposed \texttt{Us-DeFake} is composed of four components. (1) Subgraph sampling module: it employs a random walk graph sampler to obtain multiple dual-layer subgraphs, to improve the model efficiency and detect fake news in large scale social networks; (2) News propagation module: it utilizes a sampling-based graph learning model, GraphSAINT \cite{Zeng2020GraphSAINT:}, to learn propagation features of news; (3) User interaction module: it embeds credibility information into user features from user interaction relation; and (4) Fake news detection module: it classifies news into real or fake based on the final user-aware news embeddings. Specifically, the graph sampling module samples multiple independent dual-layer subgraphs, and the subsequent training process of news propagation module and user interaction module is performed in the subgraphs with bias elimination. While training these two modules, \texttt{Us-DeFake} fuses the credibility features of users into news features through the inter-layer in the dual graph to update the user-aware news embeddings for the fake news detection module. The main contributions of this work are as follows: 
\begin{itemize}
    \item \textbf{Integrality}: We propose a novel model \texttt{Us-DeFake} that mines integrated multi-relations in online social networks and learns rich user-aware information of news for accurate fake news detection.  
    \item \textbf{Effectiveness}: We illustrate the efficacy of user credibility information in user interaction relation for fake news detection. After considering this relation, the performance of the proposed \texttt{Us-DeFake} on two real-world datasets both surpasses the seven state-of-the-art baselines. 
    \item \textbf{Efficiency}: Graph sampling procedure in our method enhances its capability of efficient fake news detection in large scale social networks with massive news and users. 
\end{itemize}

\section{Related Work} \label{sec:relatedwork}
\subsection{Text-based Methods}
Text-based methods detect fake news through news content. They capture lexical and syntactic features or specific writing styles that are commonly present in fake news content. For example, P{\'e}rez-Rosas \textit{et al.} \cite{perez-rosas-etal-2018-automatic} proposed a text-driven method to detect fake news via linguistic features of news. FakeBERT \cite{kaliyar2021fakebert} detects fake news by integrating three parallel blocks of 1d-CNN into BERT \cite{devlin2018bert}. 

To further improve the performance, researchers integrated side textual information into fake news detection. dEFEND \cite{defend19shu} exploits news and the relevant comments to jointly capture the explainable sentences for discovering why the news is identified as fake or real. STANKER \cite{rao-etal-2021-stanker} takes comments as auxiliary features and adopts level-grained attention-masked BERT to distinguish the news. SRLF \cite{Yuan2021SRLFAS} is a stance-aware reinforcement learning framework that detects fake news by selecting labeled stance data for model training. Zhang \textit{et al.} \cite{emotion2021zhang} considered emotional signals in news contents, and proposed dual emotion features to represent dual emotion and the relationship between them for fake news detection. 

Albeit these text-based methods also use auxiliary features to detect fake news, they classify each piece of news independently, ignore the inherent spreading mechanism of news, and thus lack the propagation information for detecting fake news. Moreover, without user information, these methods lack the social context in fake news detection, which leads to limited results. 

\begin{figure*}[!htbp]
\centering
\centerline{\includegraphics[scale=0.14]{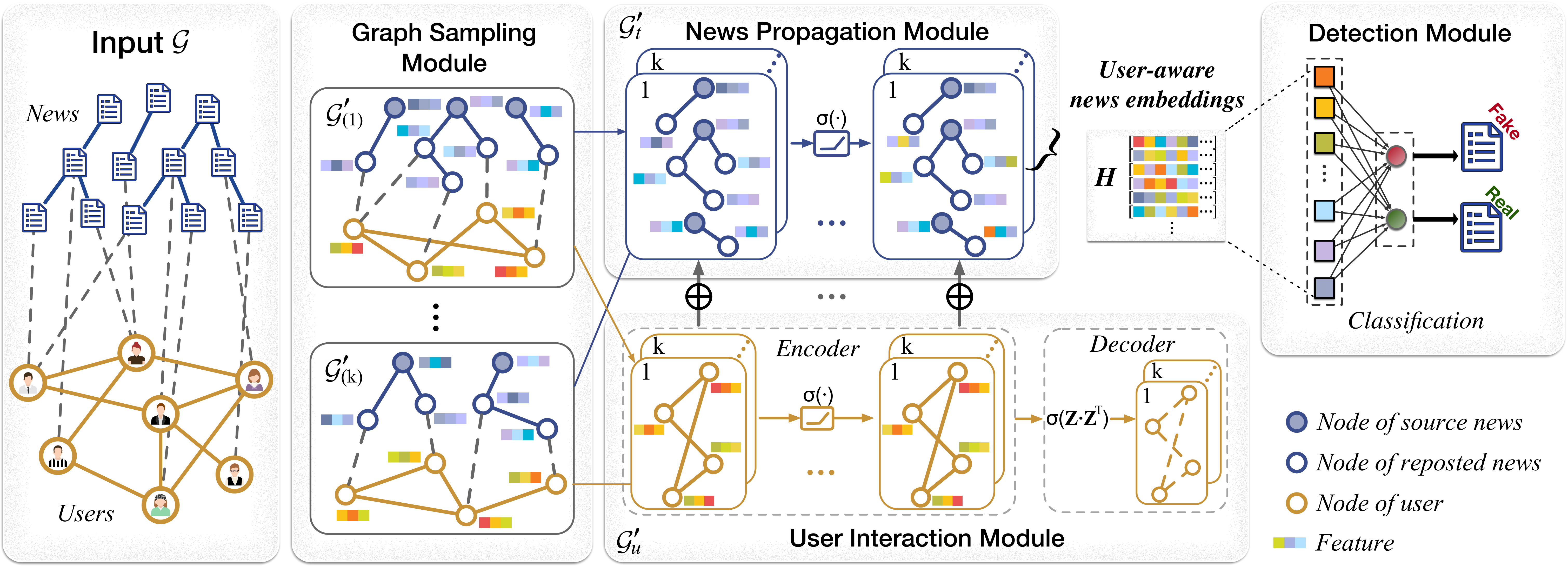}} 
\caption{The framework of \texttt{Us-DeFake} for fake news detection. \texttt{Us-DeFake} takes news and users as input in a dual-layer graph, which contains a news propagation graph and a user interaction graph as its two intra-layers, and the posting relation between news and users as its inter-layer (dotted line). Firstly, the graph sampling module samples $k$ dual-layer subgraphs $\mathcal{G}'$ with news layers and user layers. Secondly, for each subgraph, the news layer $\mathcal{G}_{t}'$ is fed into the news propagation module which adopts GCN to learn news embeddings $\bm{H}$ in a supervised setting. Meanwhile, the user layer $\mathcal{G}_{u}'$ is fed into the user interaction module that employs Autoencoder to learn user embeddings $\bm{Z}$ in an unsupervised setting. \texttt{Us-DeFake} fuses user embeddings to news embeddings through the element-wise addition operation $\oplus$ while training. The user-aware news embeddings $\bm{H}$ are obtained through the above process. Finally, the detection module regards fake news detection as a binary node classification task, and classifies the nodes of news as real or fake. Since the source news of the datasets in this work comes from fact-checking websites, the nodes of source news have no corresponding user relations.}
\label{fig:framework}
\end{figure*}

\subsection{Graph-based Methods}
With the development of graph neural networks (GNNs) \cite{wu2020comprehensive} which have a strong capability to extract information from structural data, graph learning has attracted considerable attention and has been widely used in many fields, such as community detection in social networks \cite{su2022comprehensive}. This technique of mining structural information can be applied to fake news detection as well \cite{grover2022public}, as it naturally conforms to the property of news propagation. For example, Ma \textit{et al.} \cite{ma-etal-2018-rumor} built a tree-structured graph of tweets and adopted two recursive neural networks \cite{socher2013recursive} to represent the news propagation tree, which demonstrates high effectiveness for fake news detection at the early stage of propagation. Bi-GCN \cite{bian2020rumor} is a bi-directional graph neural network model for rumor detection, it leverages GCN \cite{kipf2017semisupervised} on directed tweet graph to learn the patterns of news propagation and news dispersion. Moreover, GCN can also capture news-level structural information by constructing graphs for news pieces \cite{ding2020more}. Thus, the graph-based method is capable of representing both structural and textural properties for fake news detection. 

To provide accurate detection, some researchers incorporated external knowledge to assist in fake news detection. DETERRENT \cite{DETERRENT2020cui} is a graph attention network (GAT) \cite{velickovic2018graph} based model that uses additional information from medical knowledge graph to detect health domain misinformation. CompareNet \cite{hu2021compare} compares the topic-enriched news with the corresponding entities from the knowledge graph to learn semantically adequate features in fake news detection. KAN \cite{dun2021kan} learns both semantic-level and knowledge-level embeddings through a knowledge-aware attention network. However, it is difficult to obtain the knowledge graph of specific domains involved in some news. Learning entity representations in a large knowledge graph is a time-consuming task as well.

To consider the impact of social context, social engagements have been widely applied in fake news detection. Heterogeneous information networks (HINs) represent diverse connections \cite{shi2016survey}. Huang \textit{et al.} \cite{huang2020heterogeneous} constructed a HIN by tweets, words and users, and proposed a meta-path based GAT framework to capture misleading news. GLAN \cite{yuan2019jointly} integrates heterogeneous information by an attention-based fusion way to preserve structural and semantic features for detecting fake news. HGAT \cite{ren2021fake} employs a hierarchical attention mechanism in a HIN of news, creators and subjects, and models fake news detection as node classification. However, how to select meaningful meta-paths or execute efficient network schema learning is still the limitation of many HIN-based methods. 

Further, to detect cross-domain fake news in online social networks, Silva \textit{et al.} \cite{silva2021embracing} proposed a multimodal model. UPFD \cite{dou2021user} utilizes the users' historical posts to summarize user preferences to advance fake news detection. Both methods regard fake news detection as graph classification, that are prone to losing global information. Since graph-level modeling of fake news detection takes each piece of news and its relevant posts or users as a graph, it lacks interaction between users or posts in the entire network. 

Hence, our method \texttt{Us-DeFake} is proposed to solve the aforementioned problems under the setting of node classification. Not only does it consider textual information but also structural information in a social context. The multiple relations of news and users are employed to learn the distinctive user-aware representations of news. Compare with the existing methods, \texttt{Us-DeFake} is trained on a series of subgraphs to enhance its adaptability in large scale social networks.

\section{Definitions and Problem Statement} \label{sec:formulation}
In this section, we first describe the relevant definitions of graph, and then formulate multiple relations for fake news detection. 

DEFINITION 1. \textbf{Graph:} Let $G = (\mathcal{V}, \mathcal{E}, \mathcal{X})$ be an attributed graph, where $\mathcal{V}$ is the node set, $\mathcal{E}$ is the edge set, and $\mathcal{X}$ is a set of attributes of the corresponding nodes. $\bm{A}$ is the adjacent matrix of graph $G$.   

DEFINITION 2. \textbf{Subgraph:} $G' = (\mathcal{V}', \mathcal{E}', \mathcal{X}')$, where $\mathcal{V}'\in\mathcal{V}$, $\mathcal{E}'\in\mathcal{E}$, respectively. $ \mathcal{X}'\in\mathcal{X}$ is the attribute set of the corresponding nodes in $\mathcal{V}'$.  

DEFINITION 3. \textbf{Dual-layer Graph:} Let a dual-layer graph be $\mathcal{G} = \{G_{1}, G_{2}, \mathcal{E}_{(1,2)}\}$, where $G_{1}$ represents the graph of layer 1 and $G_{2}$ represents the graph of layer 2. $\mathcal{E}_{(1,2)}$ is the set of inter edges between $G_{1}$ and $G_{2}$ which donates the relation across the two layers. 

In this work, to model multi-relations in online social networks for fake news detection, we use a dual-layer graph to represent news, users, and the relations within or between them. A dual-layer graph with a news propagation graph and a user interaction graph can be defined as: $\mathcal{G} = (G_{t}, G_{u}, \mathcal{E}_{(t,u)})$, where $G_{t} = (\mathcal{V}_{t}, \mathcal{E}_{t}, \mathcal{X}_{t})$ is the news propagation graph, $G_{u} = (\mathcal{V}_{u}, \mathcal{E}_{u}, \mathcal{X}_{u})$ is the user interaction graph, and $\mathcal{E}_{(t,u)}$ represents the posting relation between news and users. Since the proposed method is based on graph sampling, a sampled dual-layer subgraph in our model is denoted as $\mathcal{G}' = (G'_{t}, G'_{u}, \mathcal{E}'_{(t,u)})$.  

Fake news detection is a binary classification problem. Given a dual-layer graph $\mathcal{G}$, this work models fake news detection as node classification, and aims to detect fake news in the news propagation graph with high accuracy. Each node in the news propagation graph $G_{t}$ is associated with the binary label $y\in\{0, 1\}$, where 0 denotes that the node is a piece of real news and 1 represents a piece of fake news. The fake news detection graph can be represented as $\mathcal{G} = (G_{t}, G_{u}, \mathcal{E}_{(t,u)}, \mathcal{Y})$, where $\mathcal{Y}$ is the label set of news.

\section{Methodology} \label{sec:framework}
In this section, we present the proposed \texttt{Us-DeFake} in detail. The fake news detection process of \texttt{Us-DeFake} is conducted in four components: a graph sampling module, a news propagation module, a user interaction module, and a fake news detection module. As shown in Figure \ref{fig:framework}, the real-world social network data with multi-relations is first constructed as a dual-layer graph to be the input of our model. Then, the graph sampling module samples a series of dual-layer subgraphs for fake news detection in large scale social networks. Subsequently, the news layer and the user layer of each dual-layer subgraph are fed into the news propagation module and the user interaction module, respectively, to learn the news embeddings and user embeddings. To obtain the informative user-aware embeddings of news, user embeddings are fused into the corresponding news embeddings through inter-layer edges during training. Finally, the fake news detection module adopts such distinctive news embeddings to classify fake and real news.

\subsection{Graph Sampling Module}
The spread of fake news in online social networks often involves a large number of user engagements and intensive news forwarding. This work extracts these multiple relations into a large dual-layer graph with a news layer and a user layer for fake news detection. However, it is inefficient to directly learn graph embeddings in large graphs. Thus, we first employ graph sampling to scale our method to detect fake news in large scale social networks. Through a graph sampler, multiple independent dual-layer subgraphs can be sampled at a small scale to speed up the subsequent training. This graph sampling module can solve the neighbor explosion problem in large scale social networks caused by the explosion of one piece of news, or the excessive interaction of a user. 
\begin{algorithm}[h]
  \SetAlgoLined
  \KwIn{$\mathcal{G}=(G_{t}, G_{u},\mathcal{E}_{(t,u)})$: dual-layer training graph; $r$: the number of roots; $h$: the random walk depth.}
  \KwOut{$\mathcal{G}'=(G'_{t}, G'_{u},\mathcal{E}'_{(t,u)})$: sampled dual-layer graph.}
  $\mathcal{V}_{t}^{root}$, $\mathcal{V}_{u}^{root}$ $\gets$ Select $r$ root nodes randomly from $\mathcal{V}_{t},\mathcal{V}_{u}$, respectively. 
  
  $\mathcal{V}'_{t}\gets\mathcal{V}_{t}^{root}$, $\mathcal{V}'_{u}\gets\mathcal{V}_{u}^{root}$
  
  \For{$i\in\mathcal{V}_{t}^{root}$}{
  $\{p\}\gets$ Sample neighbor nodes of $i$ in depth $h$.
  
  $\mathcal{V}'_{t}\gets\mathcal{V}'_{t}\cup\{p\}$}
  \For{$j\in\mathcal{V}_{u}^{root}$}{
  $\{q\}\gets$ Sample neighbor nodes of $j$ in depth $h$.
  
  $\mathcal{V}'_{u}\gets\mathcal{V}'_{u}\cup\{q\}$
  }
  $G'_{t}$, $G'_{u}\gets$ Subgraph of $\mathcal{G}$ includes $\mathcal{V}'_{t}$, $\mathcal{V}'_{u}$, respectively. 
  \caption{The random walk graph sampler of \texttt{Us-DeFake}} \label{alg:sampler}
\end{algorithm}

Random walk as a stochastic process has been employed in many graph samplers \cite{li2015random,pandey2020c}. In this work, we employ a random walk-based graph sampler in \cite{Zeng2020GraphSAINT:} to obtain subgraphs for data training. To sample a subgraph, we select $r$ root nodes uniformly and randomly on each intra-layer. Each walker walks on an intra-layer in $h$ hops. We sample $k$ subgraphs, where each subgraph is a dual-layer graph. The detailed graph sampling process based on a random walk is described in the Algorithm \ref{alg:sampler}. 

\subsection{News Propagation Module}
News in online social networks contains not only inherent textual information, but structural information due to sending, forwarding, and commenting on related posts. Thus, we take the textual features learned by RoBERTa \cite{liu2020roberta} as attributes to construct an attributed graph of news propagation. To efficiently learn news embeddings in the attributed news propagation graph, the training process conducts on the news subgraphs obtained by the graph sampler. Here we use a sampling based GCN model, \textit{i.e.}, GraphSAINT \cite{Zeng2020GraphSAINT:}, instead of the vanilla one for scaling \texttt{Us-DeFake} in large social networks. Considering a node $i$ in the $\ell$-th embedding layer and a node $v$ in the $\left(\ell-1\right)$-th embedding layer. If $i$ is sampled (i.e., $i \in \mathcal{V}'_{t}$), the aggregated embedding of $i$ can be calculated as:
\begin{equation}
\begin{aligned}
\bm{h}_{i}^{\left(\ell\right)} &= \sum_{v\in\mathcal{V}'_{t}}\frac{\tilde{\bm{A}}_{i,v}}{\alpha_{v,i}}\left(\bm{W}^{\left(\ell-1\right)}\right)^{T}\bm{h}_{v}^{\left(\ell-1\right)}\mathbbm{1}_{v|i} \\ 
&= \sum_{v\in\mathcal{V}'_{t}}\frac{\tilde{\bm{A}}_{i,v}}{\alpha_{v,i}}\tilde{\bm{h}}_{v}^{\left(\ell-1\right)}\mathbbm{1}_{v|i},
\end{aligned} \label{eq:aggr}
\end{equation}
where $\tilde{\bm{A}}$ is the normalized adjacency matrix, $\bm{W}$ is the weight matrix. $\bm{h}_{i}^{(0)}=\bm{x}_{i}$, and $\tilde{\bm{h}}_{v}^{\left(\ell-1\right)}=\left(\bm{W}^{\left(\ell-1\right)}\right)^{T}\bm{h}_{v}^{\left(\ell-1\right)}$. $\mathbbm{1}_{v|i}\in\{0, 1\}$ is the function to indicate whether the given $i$ is in the news propagation subgraph. $\mathbbm{1}_{v|i}=0$ if $i\in\mathcal{V}'_{t} \wedge \left(v, i\right) \notin\mathcal{E}'_{t}$; $\mathbbm{1}_{v|i}=1$ if $i\in\mathcal{V}'_{t} \wedge \left(v, i\right) \in\mathcal{E}'_{t}$; $\mathbbm{1}_{v|i}$ is not defined if $i\notin\mathcal{V}'_{t}$. $\alpha_{v, i} = \frac{p_{v, i}}{p_{i}}$ is an aggregator normalization, where $p_{v, i} = p_{i, v}$ is the probability of an edge $\left(v, i\right) \in\mathcal{E}'_{t}$  being sampled in a subgraph, and $p_{i}$ is the probability of a node $i\in\mathcal{V}'_{t}$ being sampled. This constant can eliminate biases in minibatch estimation for subgraphs. Furthermore, the other constant for bias elimination among subgraphs is loss normalization $\lambda$, which is introduced in the loss function by 
\begin{equation}
\mathcal{L}_{i}^{t} = \sum_{i\in G'_{t}}\frac{\mathcal{L}_{CE}}{\lambda_{i}}, 
\end{equation}
where $\lambda_{i} = |\mathcal{V}_{t}|\cdot p_{i}$, $\mathcal{L}_{CE}$ is the cross-entropy loss on node $i$ in the output layer. Thus, the normalized loss function of node $i$ in the news propagation graph is: 
\begin{equation}
\mathcal{L}_{i}^{t} = \sum_{i\in G'_{t}}\frac{ -y\log\hat{y}-(1-y)\log(1-\hat{y})}{\lambda_{i}},
\label{eq:loss_t}
\end{equation}
where $\hat{y}$ is the predicted value which indicates the probability of news being fake, $y\in\{0, 1\}$ is the ground truth label.

\subsection{User Interaction Module}
The creation and propagation of news are user-initiated. Thus, considering only information of the news itself but ignoring user information can lead to a lack of crucial global information. In general, trustworthy users tend to spread reliable news, while untrustworthy users are prone to propagate fake news. Regarding user credibility, besides user attributes such as verification or the number of issued Tweets, another influencing factor is user topology which reflects the interaction between users. For instance, trusted normal users commonly follow and share the news with each other, rather than untrusted zombies or virtual users. Therefore, to learn better news embeddings with distinguishing information, this work constructs a user interaction graph to capture user credibility signals, and integrates these signals into news embeddings to enhance their performance in fake news detection. 

In this module, the user interaction graph is an attributed graph, in which the learning process is in an unsupervised setting. Hence, we utilize Autoencoder \cite{hinton2006reducing}, consisting of an encoder and a decoder, to learn the user embeddings. Specifically, we adopt the sampling-based GCN as the encoder to encode the latent features of users, where the aggregation rule is calculated by Eq. (\ref{eq:aggr}) as well. To learn credibility information in user interaction from structural perspective, we employ a structural decoder where the learned latent user embeddings $\bm{Z}$ are decoded as input for the reconstruction of the original user interaction graph: 
\begin{equation}
\hat{\bm{A}} = \text{sigmoid}\left(\bm{Z}\bm{Z}^{T}\right), 
\label{eq:recons}
\end{equation}
where $\hat{\bm{A}}$ is the reconstructed user interaction graph. The learning process is guided by the objective function $\mathcal{L} = \parallel\bm{A} - \bm{\hat{A}}\parallel^{2}$. Through the bias estimation for user interaction subgraphs, a user node $j$'s normalized loss becomes 
\begin{equation}
\mathcal{L}_{j}^{u} = \sum_{j\in G'_{u}}\frac{\parallel\bm{a}_{j} - \bm{\hat{a}}_{j}\parallel^{2}}{\lambda_{j}},  
\label{eq:loss_u}
\end{equation}
where $\bm{a}_{j}\in\bm{A}$ and $\hat{\bm{a}}_{j}\in\hat{\bm{A}}$ are adjacent vector and reconstructed vector of user $j$, respectively. $\lambda_{j}$ is the loss normalization of user $j$.

\subsection{Fake News Detection Module}
In order to learn the user-aware news embeddings with rich multi-relation information for fake news detection, \texttt{Us-DeFake} fuses user embeddings into news embeddings through the inter-layer of the dual-layer graph while training the news propagation module and the user interaction module, to keep finetuning the news embeddings. Suppose the user $j$ published a piece of news $i$, the updated user-aware news embedding is calculated by
\begin{equation}
\bm{h}_{i} = \bm{h}_{i}\oplus\bm{z}_{j},
\label{eq:combine}
\end{equation}
where $\oplus$ denotes the element-wise addition operation, the news embedding vector $h_{i}$ and the user embedding vector $z_{j}$ require the same dimensions. The detailed training process of \texttt{Us-DeFake} is described in Algorithm \ref{alg:framework}. Finally, the learned user-aware news embeddings are fed to a softmax layer for fake news detection:
\begin{equation}
\hat{y} = \text{Softmax}\left(\bm{W}_{f}\bm{h}_{i}+\bm{b}_{f}\right),
\end{equation}
where $\bm{W}_{f}$ and $\bm{b}_{f}$ are the parameter matrix and vector of a linear transformation. $\hat{y}$ is the predicted value of news $i$ which indicates the probability of news being fake. The total loss function of \texttt{Us-DeFake} to minimize is as follows:
\begin{equation}
\mathcal{L} = \mathcal{L}^{t} + \mathcal{L}^{u}.  
\label{eq:totalloss}
\end{equation}

\begin{algorithm}[!htb]
  \SetAlgoLined
  \KwIn{$\mathcal{G}=(G_{t}, G_{u},\mathcal{E}_{(t,u)})$: dual-layer training graph; $\mathcal{Y}$: labels of nodes in $G_{t}$; $r, h$: parameters of graph sampling.}
  \KwOut{User-aware embeddings of nodes in $G_{t}$.}
  
  // Initialization:
  Graph sampling $\mathcal{G}'=(G'_{t}, G'_{u},\mathcal{E}'_{(t,u)})$ according to Algorithm \ref{alg:sampler}; Calculate normalization coefficients $\alpha$, $\lambda$ for $G'_{t}$, $G'_{u}$, respectively. 
  
  \For{each epoch}{
  \For{each minibatch}{
  // GCN construction for $G'_{t}=(\mathcal{V}'_{t},\mathcal{E}'_{t},\mathcal{X}'_{t})$: 
  
  Obtain $\alpha$--normalized $\bm{h}_{i}$ for news node $i$ by Eq. (\ref{eq:aggr});  
  
  // Autoencoder construction for $G'_{u}=(\mathcal{V}'_{u},\mathcal{E}'_{u},\mathcal{X}'_{t})$: 
  
  Obtain $\alpha$--normalized $\bm{z}_{j}$ for user node $j$ by Eq. (\ref{eq:recons});
  
  Update $\bm{h}_{i}$ by relation $\mathcal{E}'_{(t,u)}$ as calculated in Eq. (\ref{eq:combine});
  
  Backward propagation via the $\lambda$-normalized loss $\mathcal{L}$ in Eq. (\ref{eq:totalloss}).
  }
  }
  \Return $\bm{h}_{i}$, $\forall i\in G_{t}$. 
  \caption{The training process of \texttt{Us-DeFake}} \label{alg:framework}
\end{algorithm}

\subsection{Time Complexity}
In this work, the time complexity of the graph sampler is $\mathcal{O}(|\mathcal{V}_{t}'| + |\mathcal{V}_{u}'|) < r\cdot h$. For the news propagation module, the time complexity is linear as $\mathcal{O}(|\mathcal{E}'_{t}|d_{t}d_{t}^{1}\cdots d_{t}^{L})$, where $d_{t}$ is the dimension of input data in a subgraph and the dimension of each GCN layer is $d_{t}^{1}\cdots d_{t}^{L}$. In the user interaction module, the time complexity of Autoencoder is $\mathcal{O}(|\mathcal{V}'_{u}|d_{u}d_{u}^{1}\cdots d_{u}^{L})$. Thus, the overall time complexity of \texttt{Us-DeFake} is $\mathcal{O}(|\mathcal{V}_{t}'| + |\mathcal{V}_{u}'| + |\mathcal{E}'_{t}|d_{t}d_{t}^{1}\cdots d_{t}^{L} + |\mathcal{V}'_{u}|d_{u}d_{u}^{1}\cdots d_{u}^{L})$.

\section{Experiments} \label{sec:experiments}
In this section, we present the experimental results to evaluate the performance of \texttt{Us-DeFake}. Specifically, we aim to investigate the following questions: 
\begin{itemize}
\item \textbf{EQ1} Does \texttt{Us-DeFake} outperform all baseline methods in fake news detection? 
\item \textbf{EQ2} Does \texttt{Us-DeFake} benefit from multiple relations? 
\item \textbf{EQ3} Can user relation explain the results of fake news detection? 
\item \textbf{EQ4} Is \texttt{Us-DeFake} sensitive to the parameters? 
\item \textbf{EQ5} How efficiently does \texttt{Us-DeFake} run on large scale online social networks? 
\end{itemize}

\subsection{Experimental Setup}
\subsubsection{Datasets}
To investigate the impact of both news propagation pattern and user interaction pattern on fake news detection, we choose the FakeNewsNet datasets \cite{shu2020fakenewsnet} that contain source news fact-checked by PolitiFact\footnote{\url{https://www.politifact.com/}} or GossipCop\footnote{\url{https://www.gossipcop.com/}}, relevant tweets, and social engagement of users. For user engagement, we crawled user attributes and user-following relations. The labels are ``real'' and ``fake''. Table \ref{tab_data} summarizes the dataset statistics, where the relation ``T-T'' represents the number of edges in the news propagation graph, ``U-U'' stands for the edges in the relevant user interaction graph, and ``U-T'' represents the relation that users post tweets. 

\begin{table}[htb]
    \centering
    \caption{The Dataset Statistics.}
    \begin{tabular}{c|cc}
    \toprule[1 pt]
    \textbf{Statistics} & \textbf{Politifact} & \textbf{Gossipcop} \\ \midrule
    \multirow{2}{*}{\textbf{Source News}} & 395 & 4047  \\
    & (R: 180 / F: 215) & (R: 2444 / F: 1603) \\ \midrule
    \textbf{Tweets and Retweets} & 366,374  & 378,289 \\ \midrule
    \textbf{Users} & 195,389 & 128,912 \\ \midrule
    \multirow{3}{*}{\textbf{Relations}} & T-T: 370,025 & T-T: 386,649 \\
     & U-T: 328,608 & U-T: 328,020 \\
     & U-U: 16,193,727 & U-U: 2,724,896 \\ 
    \bottomrule[1pt]
    \end{tabular}
    \label{tab_data}
\end{table}

\begin{table*}[!t]
\centering
\small
\renewcommand\arraystretch{1}
\setlength{\tabcolsep}{1.5mm}
\caption{Overall performance for fake news detection of different methods.}
\begin{tabular}{cccccccccc} 
\toprule[1pt]
\multirow{2}{*}{\textbf{Methods}} & \multicolumn{4}{c}{\textbf{Politifact}} & \multicolumn{4}{c}{\textbf{Gossipcop}}  \\ \cmidrule(lr){2-5}\cmidrule(lr){6-9}
 & \textbf{Acc} & \textbf{Pre} & \textbf{Rec} & \textbf{F1} & \textbf{Acc} & \textbf{Pre} & \textbf{Rec} & \textbf{F1} \\ \midrule
\textbf{TextCNN} & $0.509\pm0.066$ & $0.518\pm0.062$ & $0.516\pm0.062$ & $0.506\pm0.065$ & $0.442\pm0.015$ & $0.489\pm0.012$ & $0.491\pm0.009$ & $0.432\pm0.014$ \\
\textbf{HAN} & $0.491\pm0.026$ & $0.508\pm0.03$ & $0.507\pm0.027$ & $0.484\pm0.027$ & $0.472\pm0.027$ & $0.513\pm0.017$ & $0.512\pm0.015$ & $0.466\pm0.03$ \\
\textbf{BERT} & $0.772\pm0.042$ & $0.823\pm0.028$ & $0.783\pm0.035$ & $0.801\pm0.026$ & $0.768\pm0.024$ & $0.767\pm0.022$ & $0.756\pm0.021$ & $0.758\pm0.022$ \\ 
\textbf{ALBERT} & $0.585\pm0.029$ & $0.603\pm0.166$ & $0.551\pm0.035$ & $0.502\pm0.069$ & $0.609\pm0.024$ & $0.619\pm0.166$ & $0.539\pm0.029$ & $0.505\pm0.075$ \\ \hline
\textbf{TextGCN} & $0.739\pm0.026$ & $0.742\pm0.017$ & $0.738\pm0.025$ & $0.733\pm0.022$ & $0.75\pm0.024$ & $0.649\pm0.122$ & $0.623\pm0.106$ & $0.627\pm0.107$ \\
\textbf{GraphSage} & $0.914\pm0.025$ & $0.906\pm0.025$ & $0.927\pm0.02$ & $0.911\pm0.025$ & $0.941\pm0.017$ & $0.934\pm0.018$ & $0.95\pm0.014$ & $0.939\pm0.017$ \\
\textbf{UPFD} & $0.829\pm0.006$ & $0.881\pm0.007$ & $0.767\pm0.014$ & $0.827\pm0.006$ & $0.95\pm0.023$ & $0.947\pm0.031$ & $0.954\pm0.016$ & $0.95\pm0.023$ \\ 
\textbf{Us-DeFake-A} & $\bm{0.979\pm0.011}$ & $\bm{0.975\pm0.013}$ & $\bm{0.981\pm0.011}$ & $\bm{0.978\pm0.012}$ & \underline{$0.954\pm0.011$} & \underline{$0.951\pm0.016$} & \underline{$0.964\pm0.009$} & \underline{$0.955\pm0.014$} \\
\textbf{Us-DeFake-C} & \underline{$0.967\pm0.03$} & \underline{$0.962\pm0.033$} & \underline{$0.973\pm0.024$} & \underline{$0.965\pm0.029$} & $\bm{0.974\pm0.013}$ & $\bm{0.97\pm0.015}$ & $\bm{0.977\pm0.012}$ & $\bm{0.973\pm0.014}$ \\
\bottomrule[1pt]
\end{tabular}
\label{tab-results}
\end{table*}

\subsubsection{Baselines}
We compare our approach with seven representative baseline algorithms: 
\begin{itemize}
\item \textbf{TextCNN} \cite{kim-2014-convolutional} builds convolutional neural networks (CNNs) on \textit{word2vec} of news contents for sentence classification, it utilizes multiple convolution filters to capture different granularity of text features. 
\item \textbf{HAN} \cite{yang2016hierarchical} is a hierarchical attention network for document classification. It encodes a two-level attention mechanism for news contents, \textit{i.e.}, sentence-level and word-level attention. 
\item \textbf{BERT} \cite{devlin2018bert} is a pre-trained language model that uses a bidirectional encoder and self-attention heads. For fake news detection, BERT can represent semantic signals in the news.  
\item \textbf{ALBERT} \cite{lan2019albert} is a lite BERT. It adds a pretraining objective while reducing the number of parameters in BERT for efficient computation on large scale data. 
\item \textbf{TextGCN} \cite{yao2019graph} extends the semi-supervised model of graph neural networks (GCN \cite{kipf2017semisupervised}) for text classification. In fake news detection, GCN can consider structural information by constructing a word graph of each piece of news. 
\item \textbf{GraphSage} \cite{NIPS2017_5dd9db5e} is a classic inductive graph representation method on large networks. In this work, we utilize text embeddings as node attributes and construct a graph of news propagation for detecting fake news. 
\item \textbf{UPFD} \cite{dou2021user} is a user preference-aware fake news detection method. It combines users' endogenous preferences by encoding historical posts, and the task of fake news detection is modeled as graph classification. 
\end{itemize}

\subsubsection{Experimental Settings and Implementation}
For all models, the train-validation-test split of two datasets is 70\%-10\%-20\%. The experimental results are averaged over 5--fold cross validation. We implement our method with Pytorch. All the experiments are conducted by Python 3.6, 1 NVIDIA Volta GPU, and 395G DRAM. 

In the experiment of \texttt{Us-DeFake}, we set the number of roots to 3000 in the graph sampler module, and the random walk depth to 2, since the news propagation graph is composed of tree structures with a depth of 3. To construct a news propagation graph, ReBERTa \cite{liu2020roberta} is used to learn 768-dimensional embeddings as node attributes. The embedding size in both the news propagation module and the user interaction module is set to 512, and the number of graph convolution layers is set to 2. For model optimization, we adopt the Adam optimizer and set the learning rate to 0.01 for both datasets. We set the number of epoch to 30. For a fair comparison with baselines that do not consider news propagation, we only evaluate the classification results of source news for all methods.   

\subsubsection{Metrics}
To evaluate the performance of approaches in this work, we adopt four commonly used metrics for evaluation, \emph{i.e.,} Accuracy (ACC), Precision (Pre), Recall (Rec), and F1 score (F1).

\subsection{Performance Evaluation (EQ1)}
To evaluate the overall performance of our method, we compare the proposed \texttt{Us-DeFake} with seven popular baselines in two datasets, and measure them by four evaluation metrics. In order to extend our method, we also use GAT \cite{velickovic2018graph} in the convolutional layer as a variant of \texttt{Us-DeFake}. Table \ref{tab-results} shows the overall experimental results, where ``\texttt{Us-DeFake-C}'' represents our method using GCN and ``\texttt{Us-DeFake-A}'' represents our method using GAT. Bold marks optimal results, underline indicates suboptimal results.

As reflected in Table \ref{tab-results}, both our method and its variant achieve optimal or suboptimal results for all metrics on two datasets, which indicates that mining multiple relations in social networks is conducive to fake news detection. Such results reveal that the user-aware news embeddings of \texttt{Us-DeFake} indeed learn sufficient information rather than other baseline methods. The reason why \texttt{Us-DeFake-A} performs better than \texttt{Us-DeFake-C} on Politifact is that users interact more frequently in this dataset. This is illustrated in Table \ref{tab_data}, where the number of U-U relations in Politifact is more than in Gossipcop. In \texttt{Us-DeFake-A}, the capability of the attention mechanism to capture more information from vital neighbors provides distinctive user signals to user-aware news embeddings, which in return improves the results of fake news detection.     

We divide the baselines into two groups, as shown in Table \ref{tab-results}, the first four are text-based methods, while the remainders are graph-based methods. The graph-based approaches integrally perform better than the text-based ones, because the graph models consider structural information as well as textual information. Both the inter-sentence or inter-word structure of news and the structure of news propagation can benefit the detection of fake news. ALBERT, as an improved algorithm based on BERT, is expected to perform better. However, as a lite model, parameter reduction affects ALBERT's performance on fake news detection. 

Among the graph-based approaches, GraphSage shows a stable performance that surpasses all the text-based methods, because we built the news propagation graph for it. The graph learns the spread property of news. The performance of UPFD comes close to our proposed model. UPFD adapts user information to enhance fake news detection as well, but it utilizes users' historical information, while we consider the users' interaction relationship. However, the gap between UPFD's performance in the two datasets is large, and the performance is not as stable as \texttt{Us-DeFake}'s. To sum up, \texttt{Us-DeFake} achieves stable and accurate results in fake news detection, and is superior to all baseline methods.

\subsection{Ablation Study (EQ2)}
\begin{figure}[!htbp]
\centering
\subfigure[Politifact]{\includegraphics[width=0.236\textwidth]{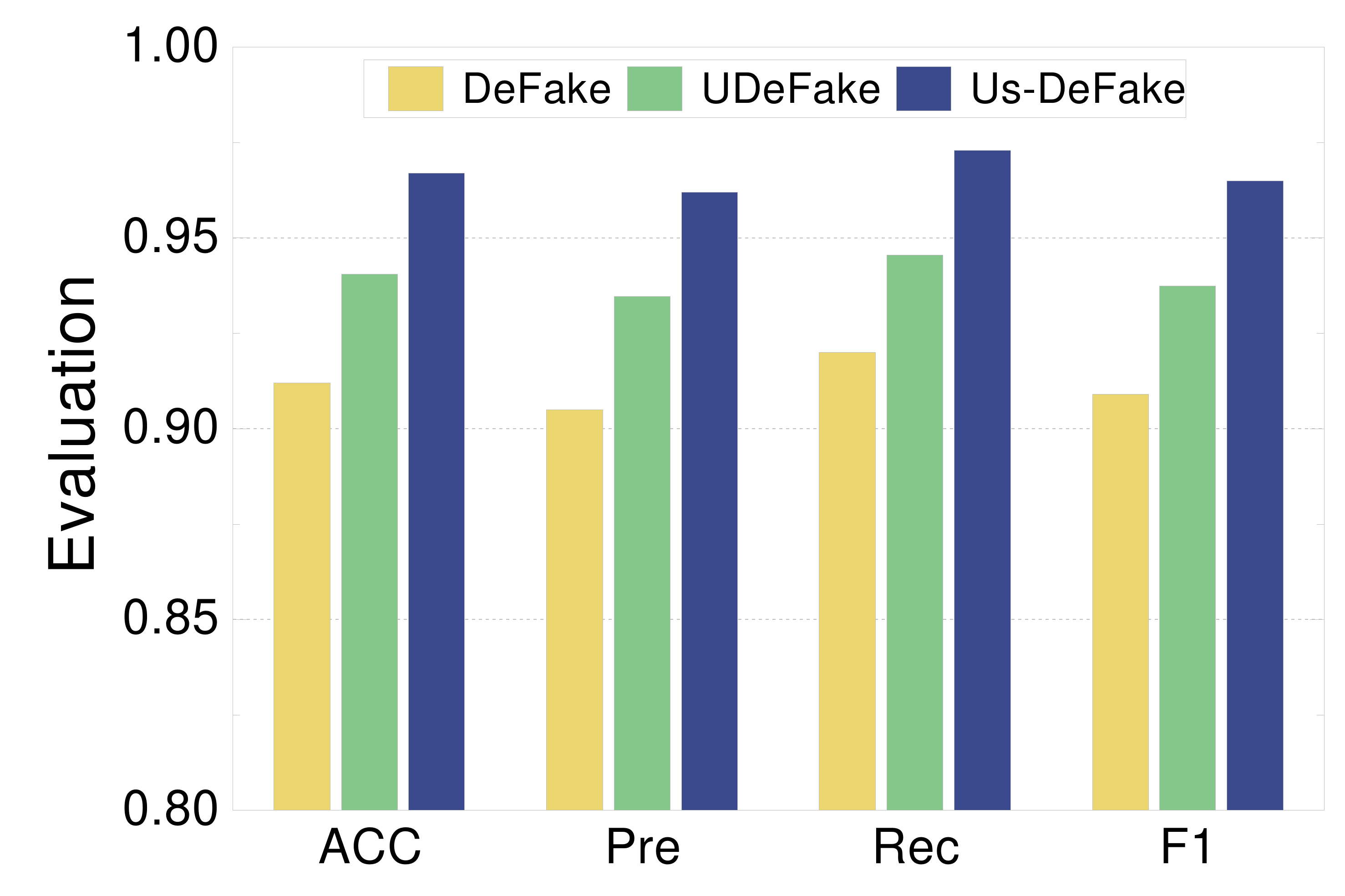}}
\subfigure[Gossipcop]{\includegraphics[width=0.236\textwidth]{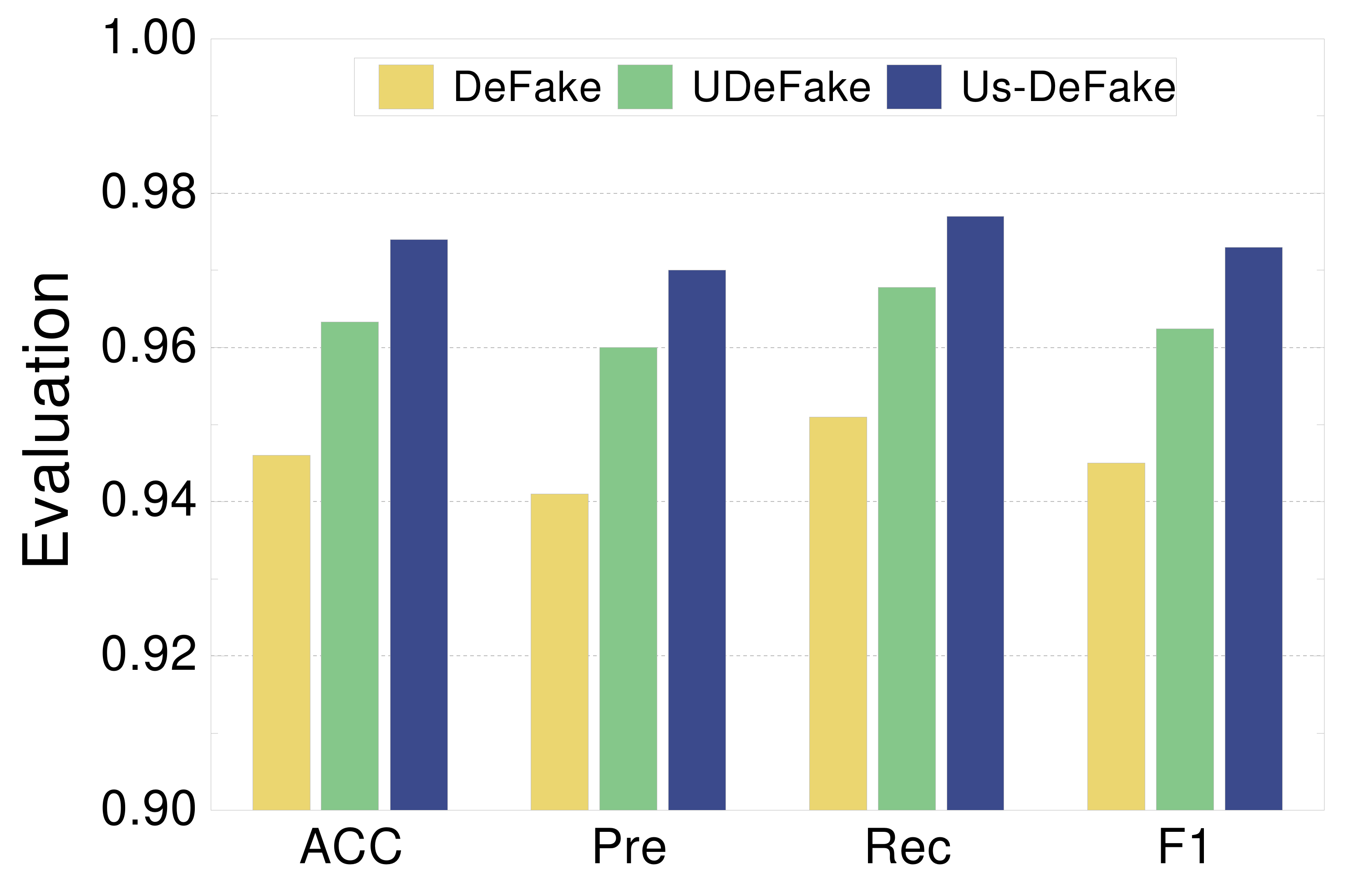}}
\caption{Ablation study.}
\label{fig:ablation}
\end{figure}

To demonstrate the effectiveness of multi-relations in \texttt{Us-DeFake} for fake news detection, we conduct ablation experiments to examine the effectiveness of the news propagation relation, the user interaction relation, and the posting relation between users and news. Figure \ref{fig:ablation} shows the experimental results of the ablation study, where ``\texttt{DeFake}'' indicates that the new propagation module is solely used without user-relevant relations. ``\texttt{UDeFake}'' is the model with the news propagation module and the user interaction module, but it fuses final user embeddings into news embeddings after separate training. ``\texttt{Us-DeFake}'' integrates user embeddings into news embeddings while training to continuously enhance the distinctiveness of news embeddings by considering the posting relation.  

All results in \texttt{DeFake} are greater than 0.9 in both datasets, which indicates that the propagation embeddings with textual features learned by the news propagation module can provide useful information for fake news detection. In \texttt{UDeFake}, after integrating the user interaction module with the news propagation module, all results increase significantly in both datasets. This phenomenon illustrates that considering user interaction relation enriches the information of user-aware news embeddings. The performance of \texttt{Us-DeFake} is further improved, with values of all evaluation metrics above 0.95. This is because fusing the user embeddings into the news embeddings through the posting relation while training can keep finetuning the news embeddings. By capturing three multi-relations, \texttt{Us-DeFake} obtains more enhanced, discriminative, and informative embeddings for fake news detection.    

\subsection{Case Study (EQ3)}
\begin{figure}[!htbp]
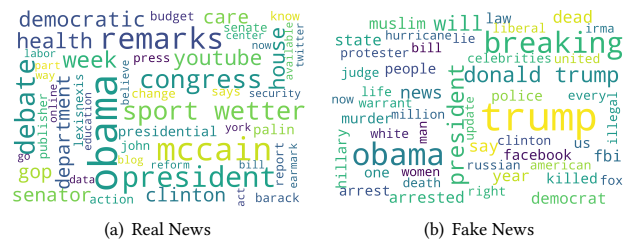

\centering
\subfigure[Real News]{\includegraphics[width=0.22\textwidth]{wordcloud_real}}
\hspace{0.2cm}
\subfigure[Fake News]{\includegraphics[width=0.22\textwidth]{wordcloud_fake}}
\caption{Word clouds of news in the Politifact dataset.}
\label{fig:wordcloud}
\end{figure}

To explore the correlation between news authenticity and user credibility, we conduct a case study on the Politifact dataset about political news. We generate word clouds for real news and fake news, respectively. As shown in Figure \ref{fig:wordcloud}, the high-frequency words that appear in real news and fake news are different. This confirms the rationality of the news propagation module, which learns the distinguishing news features of real news and fake news in a news propagation graph. 

\begin{figure*}[!tb]
\centering
\subfigure[]{\includegraphics[width=0.24\textwidth]{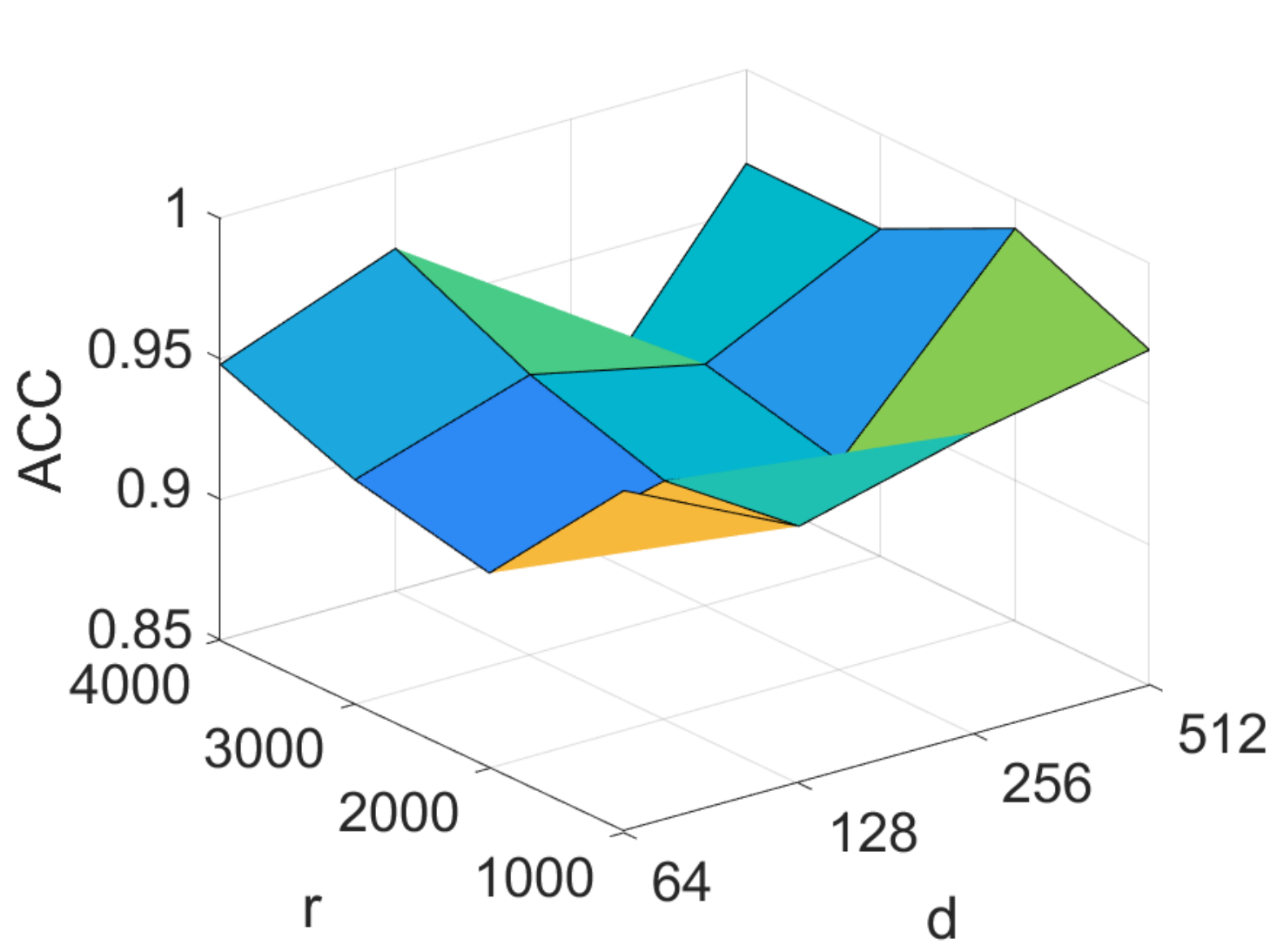}}
\hspace{0.2cm}
\subfigure[]{\includegraphics[width=0.24\textwidth]{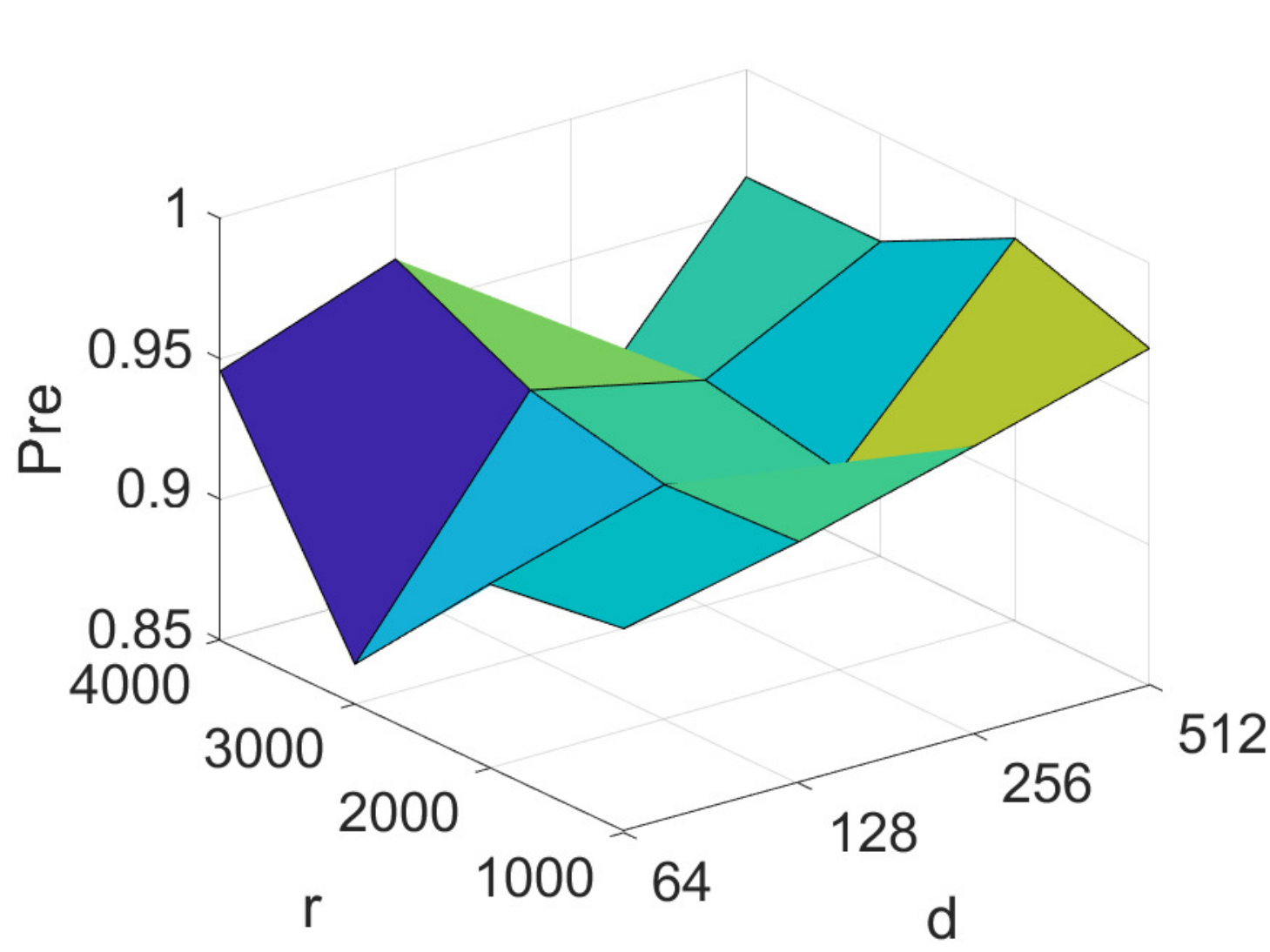}}
\subfigure[]{\includegraphics[width=0.24\textwidth]{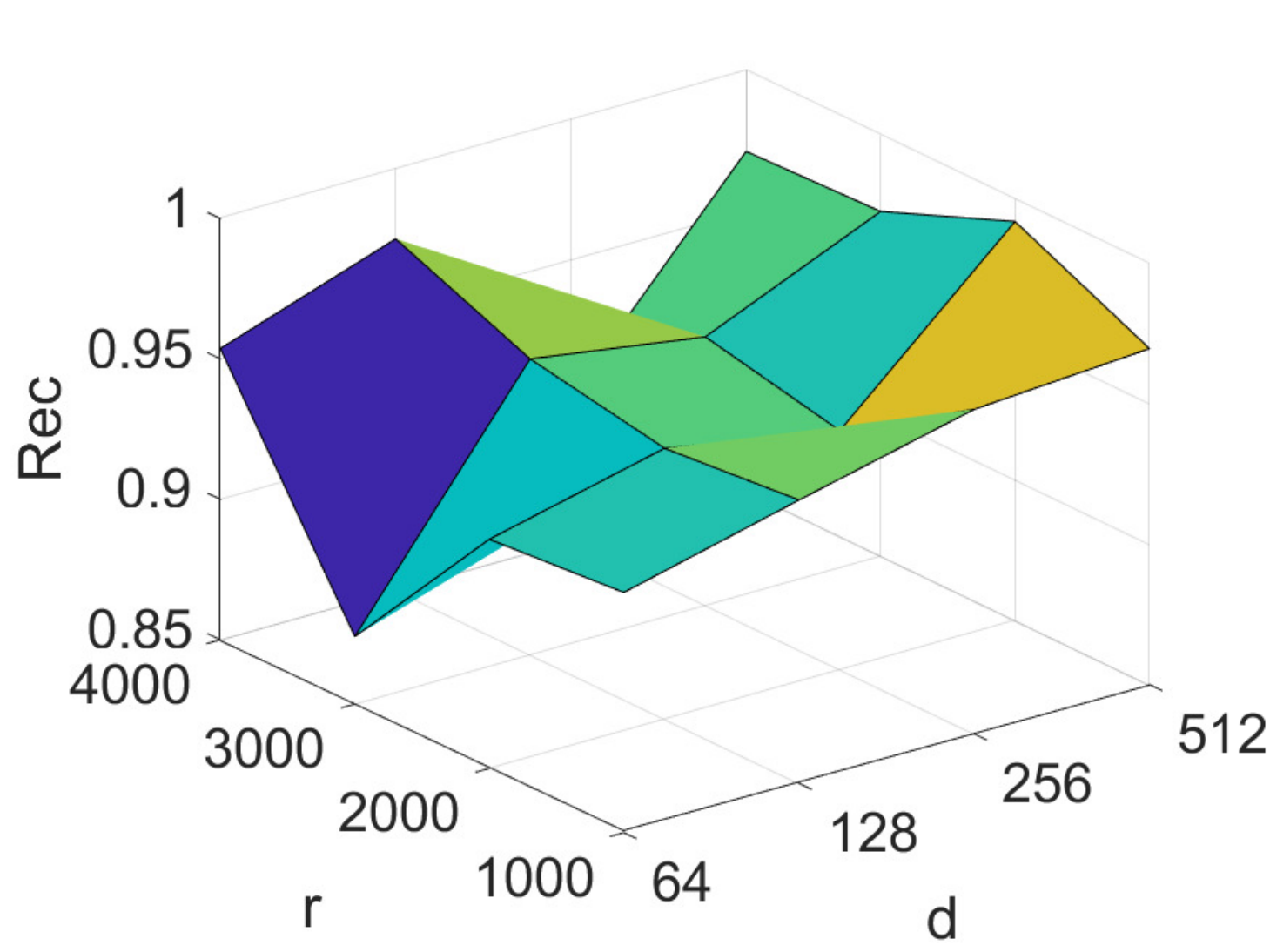}}
\subfigure[]{\includegraphics[width=0.24\textwidth]{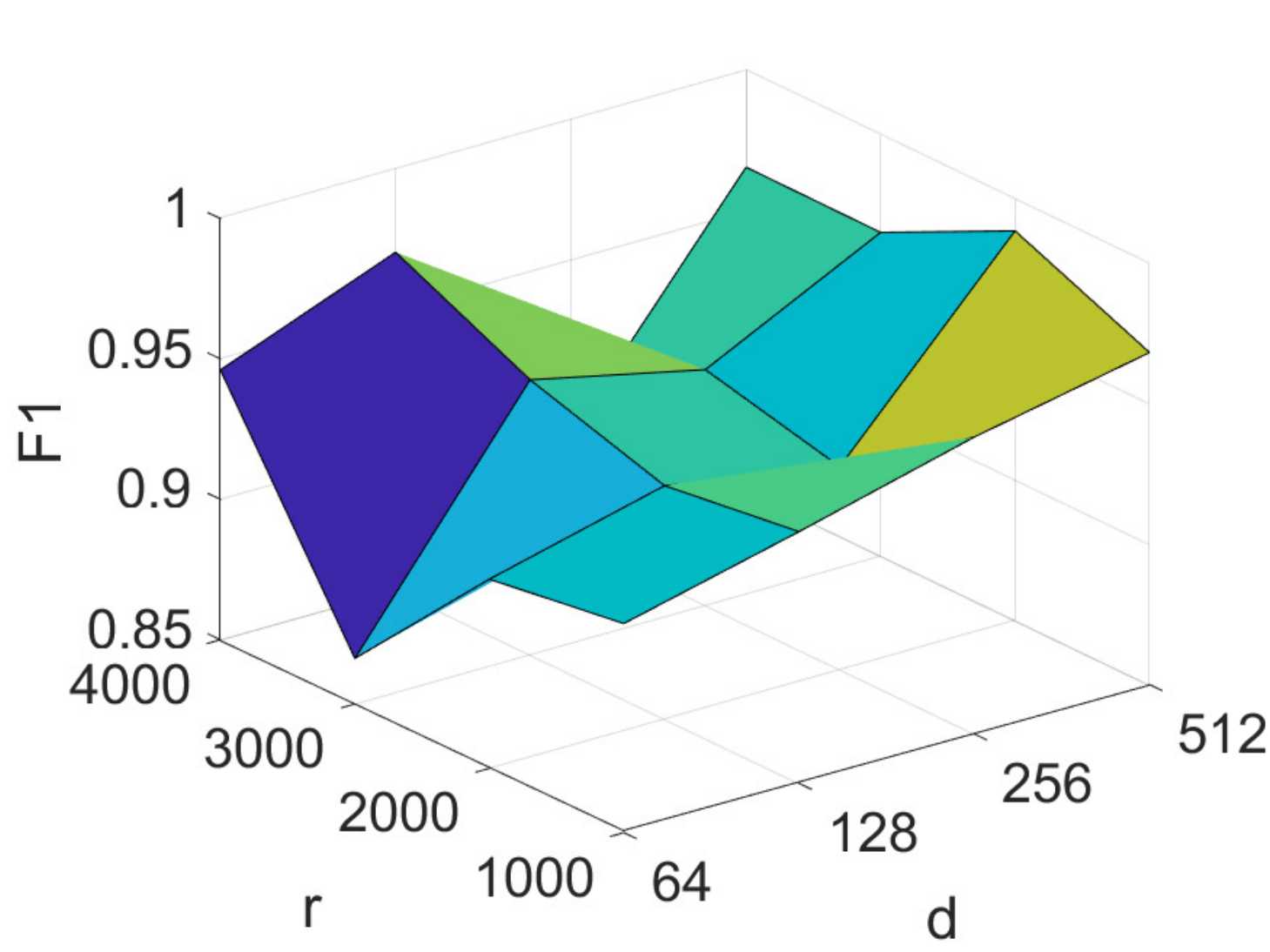}}
\subfigure[]{\includegraphics[width=0.24\textwidth]{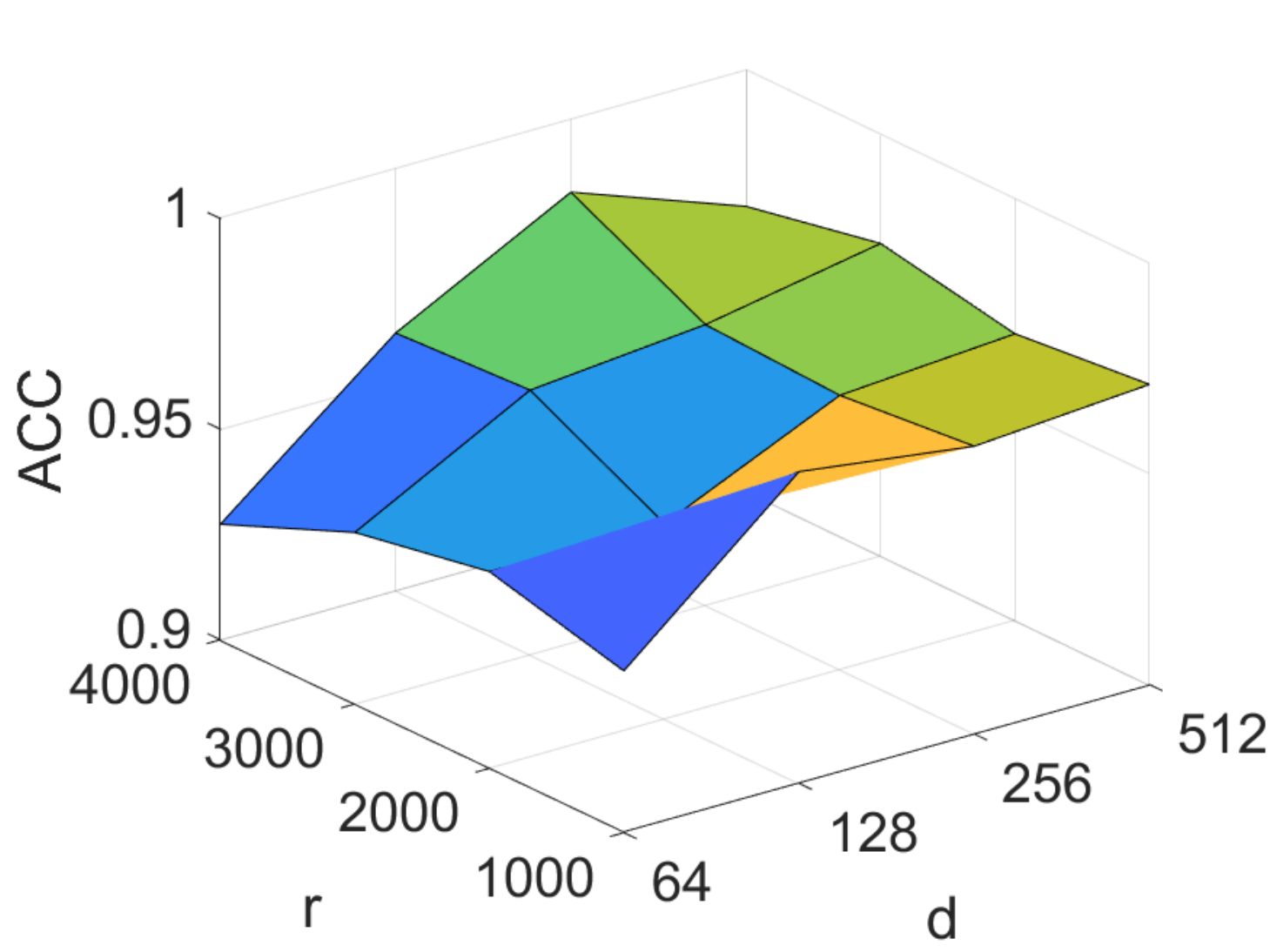}}
\hspace{0.2cm}
\subfigure[]{\includegraphics[width=0.24\textwidth]{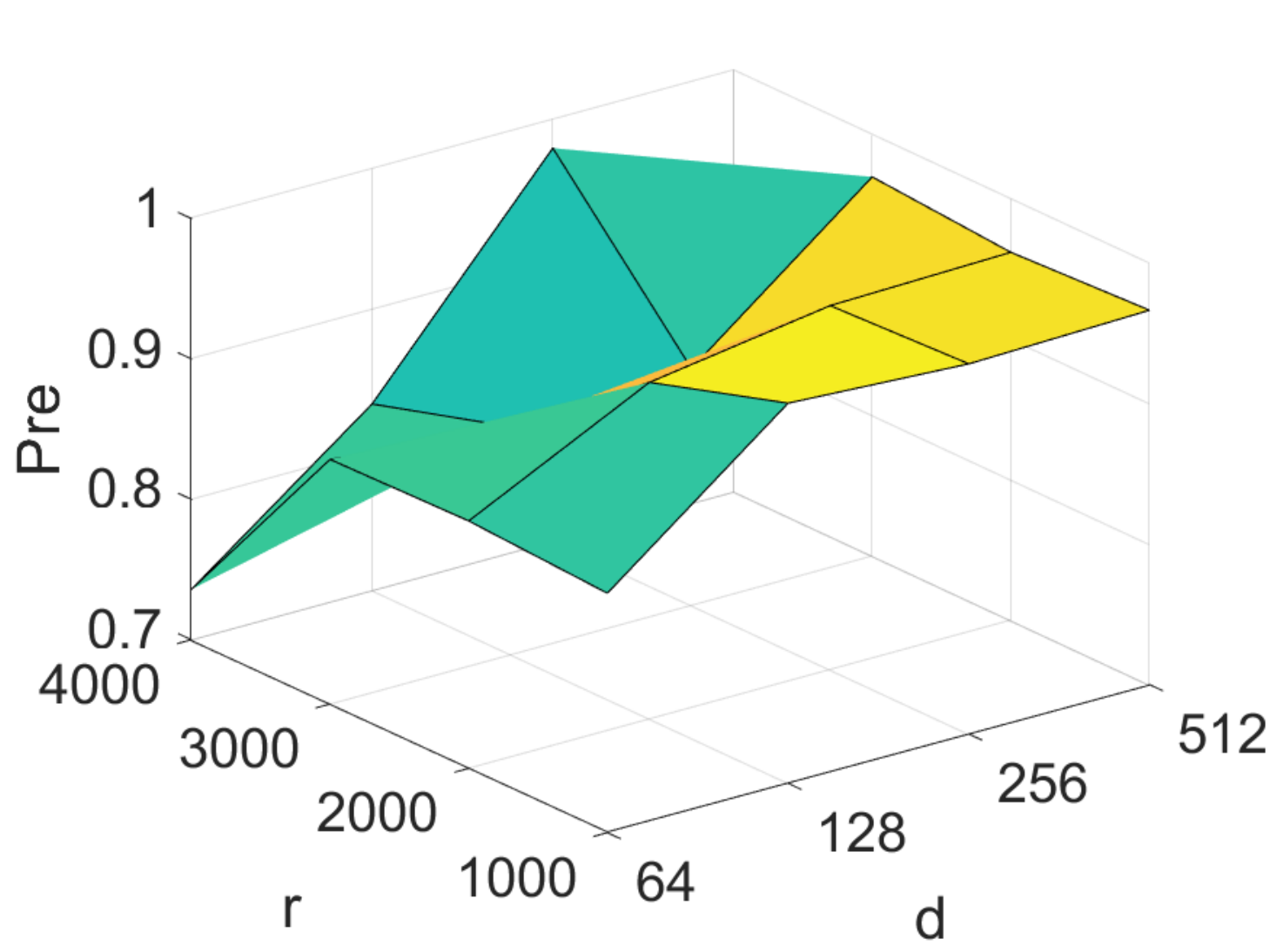}}
\subfigure[]{\includegraphics[width=0.24\textwidth]{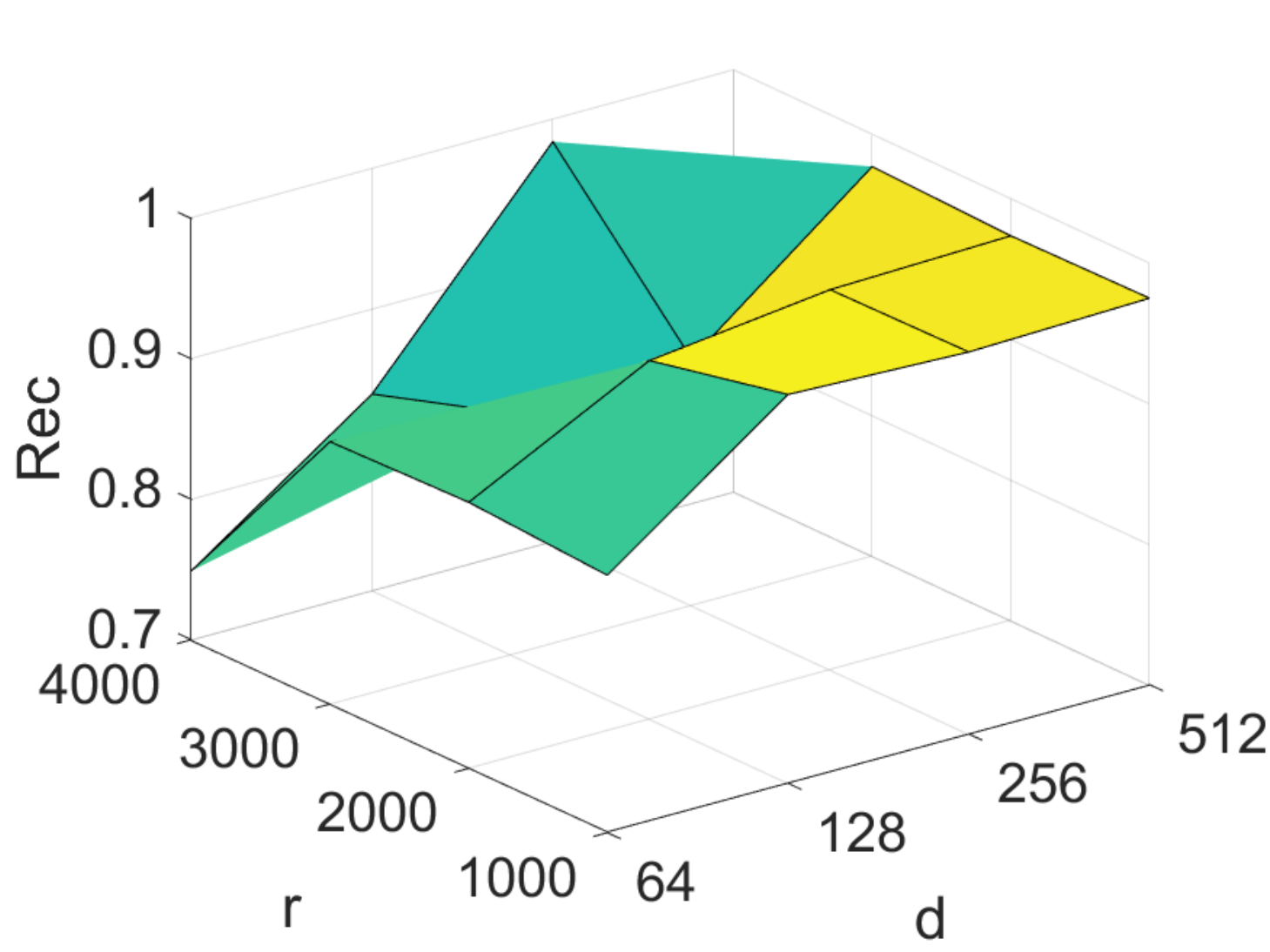}}
\subfigure[]{\includegraphics[width=0.24\textwidth]{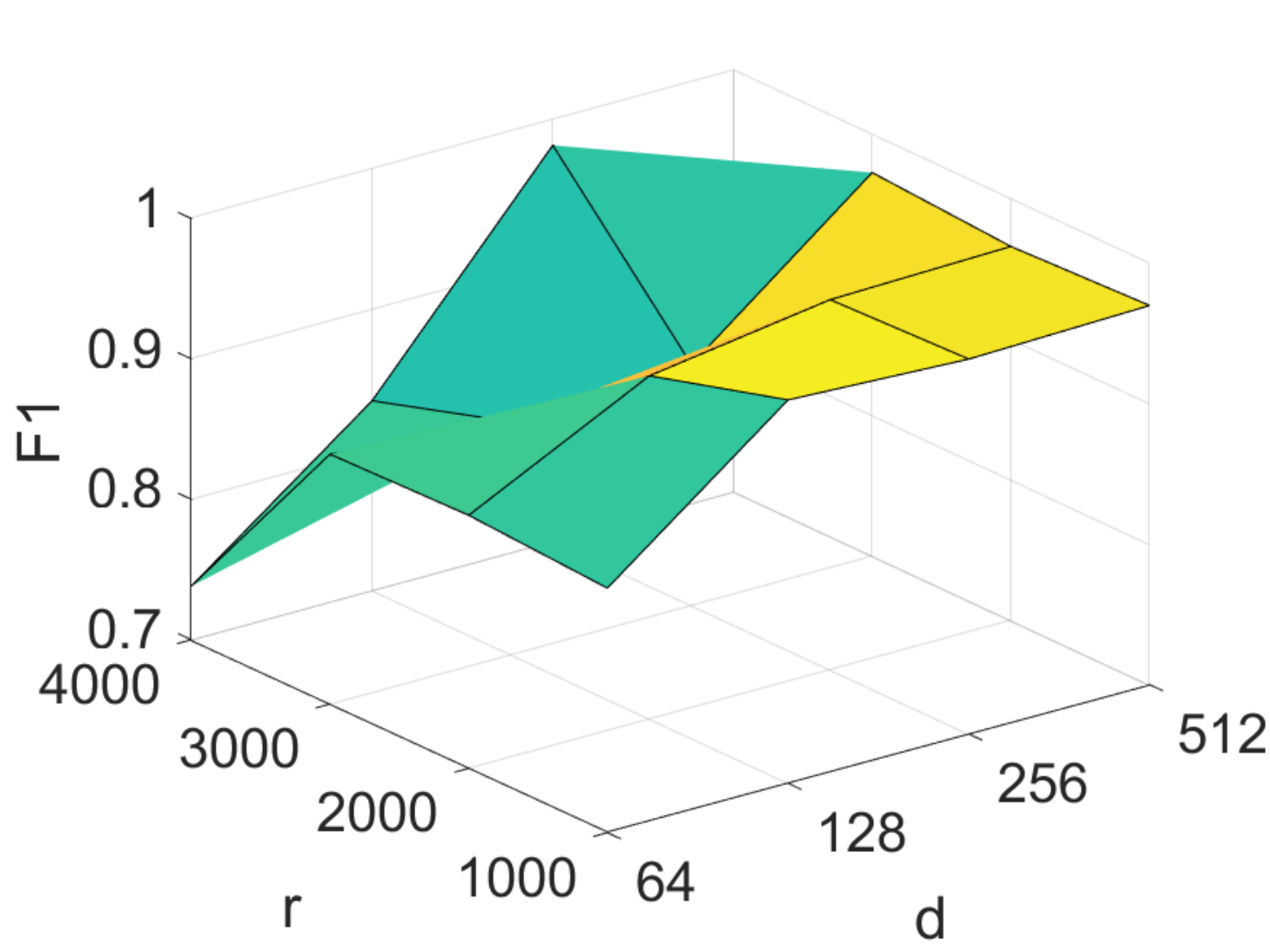}}
\caption{Parameter analysis of the proposed \texttt{Us-DeFake}. (a)--(d) on Politifact dataset, (e)--(h) on Gossipcop dataset. }
\label{fig:para_analysis}
\end{figure*}

Further, to reveal the correlation between news and users, we randomly choose attributes of four relevant users for analysis. The main attributes of these users are shown in Table \ref{tab:user_attr}, where User1 and User2 published real news, User3 and User4 published fake news. The columns of ``Friends'' (\emph{i.e.}, following) and ``Followers'' represent structural information in the user interaction graph as well. According to Table \ref{tab:user_attr}, User1 is a verified public account with lots of followers, so it has a strong influence and is reliable in news spreading. User2 is supposed to be an active normal user with a close number of friends and followers. ``Listed'' indicates the number of public lists of which this user is a member. Thus, User2 is also trustworthy, because a normal user is not likely to intentionally spread fake news to his or her friends. 

\begin{table}[!htb]
    \centering
    \caption{Part of user attributes of four randomly-selected users for a case study. }
    \begin{tabular}{c|ccccc}
    \toprule[1 pt]
    \textbf{User} & \textbf{Friends} & \textbf{Followers} & \textbf{Status} & \textbf{Listed} & \textbf{Verified} \\ \midrule
    User1 & 870 & 51,693,630 & 463,743 & 213,104 & $\checkmark$ \\
    User2 & 6775 & 7579 & 138197 & 608 & $\times$ \\ \hline
    User3 & 2388 & 210 & 71540 & 12 & $\times$ \\
    User4 & 94 & 0 & 327 & 0 & $\times$ \\
    \bottomrule[1pt]
    \end{tabular}
    \label{tab:user_attr}
\end{table}
User3 and User4 have a low ranking of trust, due to the disproportionate number of friends and followers, or a large amount of ``Status'' with rarely ``Listed''. These phenomena confirm that news published by reliable users is usually trustworthy, while fake news is generally spread by unreliable users. It is reasonable to employ a user interaction module to assist fake news detection in our method. User reliability can explain why the news is detected as real or fake.

\subsection{Parameter Analysis (EQ4)}
We also investigated the sensitivity of \texttt{Us-DeFake} to the parameters in regard to the number of roots $r$ when sampling subgraphs and different embedding sizes $d$ in the news propagation module and the user interaction module. Figure \ref{fig:para_analysis} shows ACC, Pre, Rec, and F1 results for datasets of Politifact and Gossipcop. The results demonstrate that the number of roots in the graph sampler does not affect results, but the results are unstable when the embedding dimension is set to 64. In general, our method maintains acceptable results in most parameter combinations steadily. 

\subsection{Time Efficiency (EQ5)}
\begin{figure}[!htbp]
\centering
\centerline{\includegraphics[scale=0.21]{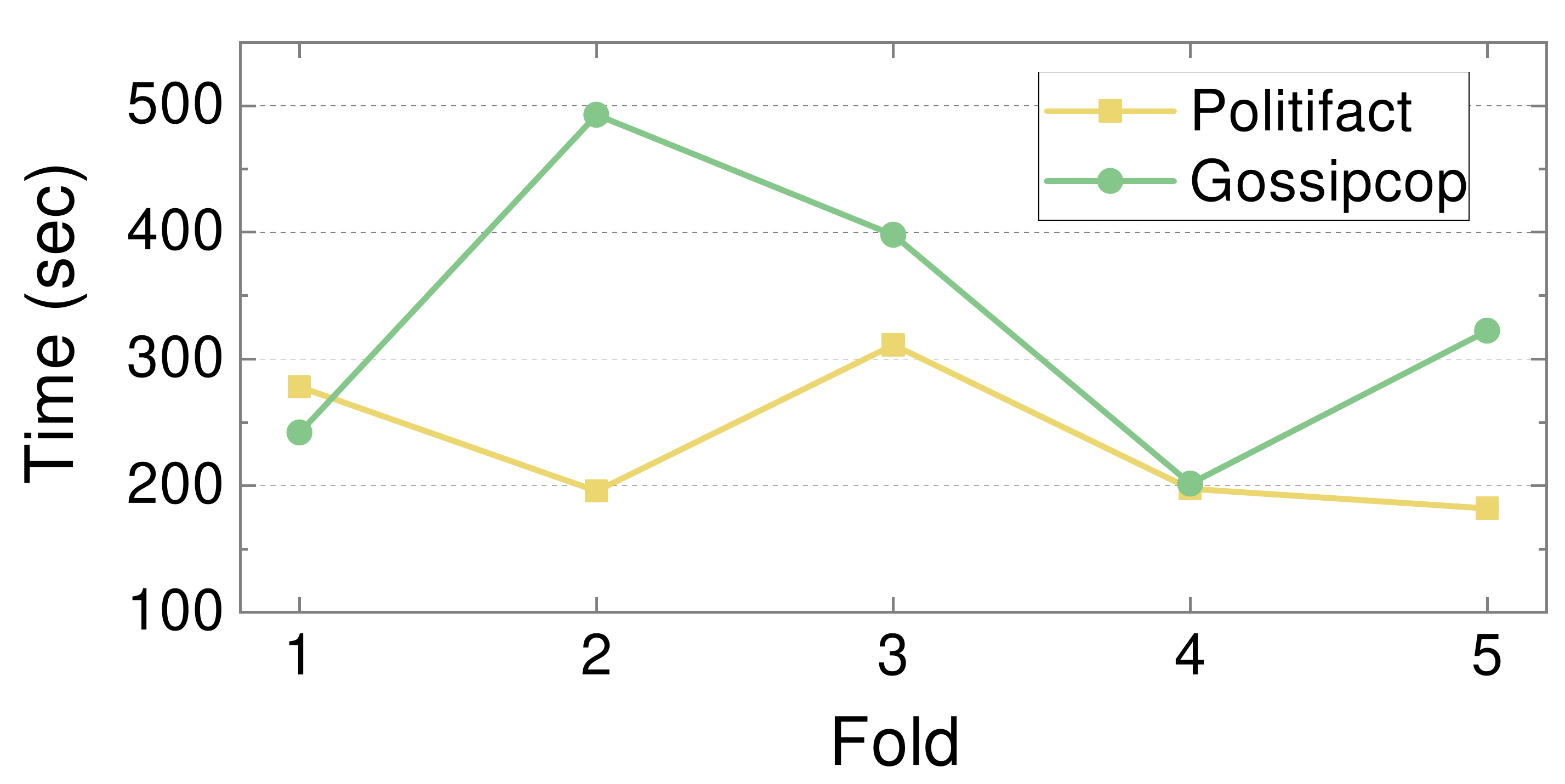}} 
\caption{Runtime of \texttt{Us-DeFake}.}
\label{fig:runtime}
\end{figure}

To evaluate \texttt{Us-DeFake}'s runtime performance in large scale social networks, we recorded its training time in each epoch. Since the input sizes of the same datasets vary in each method (\emph{e.g.}, the user interaction graph in \texttt{Us-DeFake} is not considered by other baselines), it is unfair to compare the runtime with each other. Figure \ref{fig:runtime} shows \texttt{Us-DeFake}'s average training time of one epoch in the datasets with the 5-fold split. To further speed up the training time and degrade the adverse impact of inveracious interaction between users, we ignore edges between two users whose Jaccard similarity is less than 0.1, so that our model can restore actual user relations and run in less time. The runtime in Figure \ref{fig:runtime} illustrates that \texttt{Us-DeFake} performs efficiently on both datasets with more than half a million nodes, but the density of the subgraphs obtained by sampling causes fluctuations in runtime. 

\section{Conclusion} \label{sec:conclusion}
In this work, we propose a user-aware model \texttt{Us-DeFake} to incorporate multi-relations between news and users for fake news detection. We uncover that the interaction relation between users reflects user credibility, and such credibility has a positive correlation with news authenticity. Based on a dual-layer graph construction for multi-relation exhibition, we first adopt a graph sampler to scale fake news detection in large scale social networks. Further, we design the modules of news propagation and user interaction to integrate distinctive user credibility signals into news embeddings for detecting fake news. The extensive experiments on large scale real-world datasets demonstrate the superiority of \texttt{Us-DeFake}, which notably outperforms seven popular baseline methods. In the future, we plan to address the problem of training time instability caused by the graph sampler in \texttt{Us-DeFake}, and explore early detection of fake news in propagation. 

\begin{acks}
This work was supported by the Australian Research Council (ARC) Projects Nos. DE200100964, LP210301259, and DP230100899. 
\end{acks}

\clearpage
\bibliographystyle{ACM-Reference-Format}
\balance
\bibliography{main}

\end{document}